\documentclass[aps,prl,twocolumn,showpacs]{revtex4}
\usepackage{graphicx}
\usepackage{color} 
\usepackage{amsmath}

\newcommand{\overbar}[1]{\mkern 1.5mu\overline{\mkern-1.5mu#1\mkern-1.5mu}\mkern 1.5mu}

\begin{document}
\title{Reconstruction-induced trefoil knot Fermi contour of Au(111)}
\author{Maciej Dendzik}
\author{Marco Bianchi}
\author{Matteo Michiardi}
\author{Charlotte E. Sanders}
\author{Philip Hofmann}
\email{philip@phys.au.dk}
\affiliation{Department of Physics and Astronomy, Interdisciplinary Nanoscience Center (iNANO), Aarhus University, 8000 Aarhus C, Denmark}
\date{\today}
\begin{abstract}
Using angle-resolved photoemission spectroscopy (ARPES), we study the effect of the so-called herringbone reconstruction of Au(111) on the dispersion of the free electron-like surface state. While earlier ARPES investigations have only reported a minor interplay of the surface state dispersion and the underlying reconstruction, we show that the uniaxial lattice distortion and the thereby changed reciprocal lattice for the first atomic layer leads to distinct surface state dispersions around the first order reciprocal lattice points of the three domains, creating a constant energy surface resembling a trefoil knot. The findings resolve the long-standing discrepancy between, on one hand, the reconstruction-induced surface state modifications reported in scanning tunnelling microscopy and first principle calculations and, on the other hand, their conspicuous absence in photoemission.  
\end{abstract}

\maketitle
\section{Introduction}
Two-dimensional superlattices and moir\'e structures have recently received renewed attention because of the possibility of using them to engineer the electronic structure of stacked two-dimensional materials. It is, for instance, possible to introduce an electronic length scale comparable to the magnetic length in a strong field, helping to create situations like the Hofstadter butterfly \cite{Hunt:2013aa,Dean:2013aa,Ponomarenko:2013aa,Woods:2014aa}. Due to the particularly important role of graphene in this type of transport phenomena, attention has mostly focused on hexagonal moir\'e structures \cite{Zeller:2014aa}, such as the one formed between (bilayer) graphene and hexagonal boron nitride (hBN)  \cite{Hunt:2013aa,Dean:2013aa,Ponomarenko:2013aa,Woods:2014aa}, Ir(111) \cite{Pletikosic:2009aa} or SiC(0001)  \cite{Bostwick:2007ac}; or, between single layer MoS$_2$ and Au(111) \cite{Miwa:2015aa,Gronborg:2015aa}. The effect of the moir\'e structure on the electronic structure of the two-dimensional material varies strongly for the different cases, depending on the orbital character of the bands involved and the interlayer interaction. For graphene on hBN  or Ir(111), for instance, the interaction leads to replicas and band gap openings near the boundaries of the superstructure Brillouin zone, but no such effects are observed for graphene on SiC or MoS$_2$ on Au(111). 

A situation closely related to the moir\'e of two hexagonal lattices already arises in the so-called herringbone reconstruction of the clean Au(111) surface \cite{Hove:1981aa,Barth:1990aa}. In this reconstruction a moir\'e-like superstructure is formed between the topmost layer of atoms, which undergoes a uniaxial compression by 4.55\%, and the second layer with a perfect hexagonal structure. This reconstructed surface hosts a free electron-like surface state \cite{Heimann77} that has been intensely studied using angle-resolved photoemission spectroscopy (ARPES) and scanning tunnelling microscopy (STM), revealing phenomena such as spin-orbit splitting of surface states \cite{LaShell:1996aa,Henk:2003aa,Hoesch:2004aa}, quasiparticle interference in the presence of spin-polarised bands \cite{Petersen:2000ab}, surface state lifetimes \cite{Kliewer:2000aa,Reinert:2001aa}, and lifetimes in spin-split systems \cite{Nechaev:2009aa}. Despite the major surface reconstruction, ARPES data show only very minor effects of the atomic arrangement on the observed surface state dispersion \cite{Reinert:2004ab} and cross section \cite{Borghetti:2012aa}. This is hard to reconcile with several indications from STM \cite{Chen:1998aa,Burgi:2002aa} and theory \cite{Takeuchi:1991aa,Seitsonen:2016aa} that the reconstruction has a non-negligible effect on the surface state wave function. 

Here we show that the surface reconstruction does indeed have a major influence on the dispersion probed by ARPES, but only on the dispersion measured away from normal emission in a surface Brillouin zone (SBZ) center corresponding to a higher diffraction order. The uniaxial compression of the lattice directly affects the reciprocal lattice and thus the position of the higher order zone centers. Together with the three rotational domains present on the surface, the measured dispersion turns from a simple parabolic state with a circular Fermi contour to a complex dispersion with constant energy surfaces resembling a trefoil knot. We emphasize that this dominating Fermi contour does not simply consist of replicas caused by the overall moir\'e periodicity but rather by displaced dispersions induced by the local lattice distortion. 

\begin{figure*}
\includegraphics[width=0.8\textwidth]{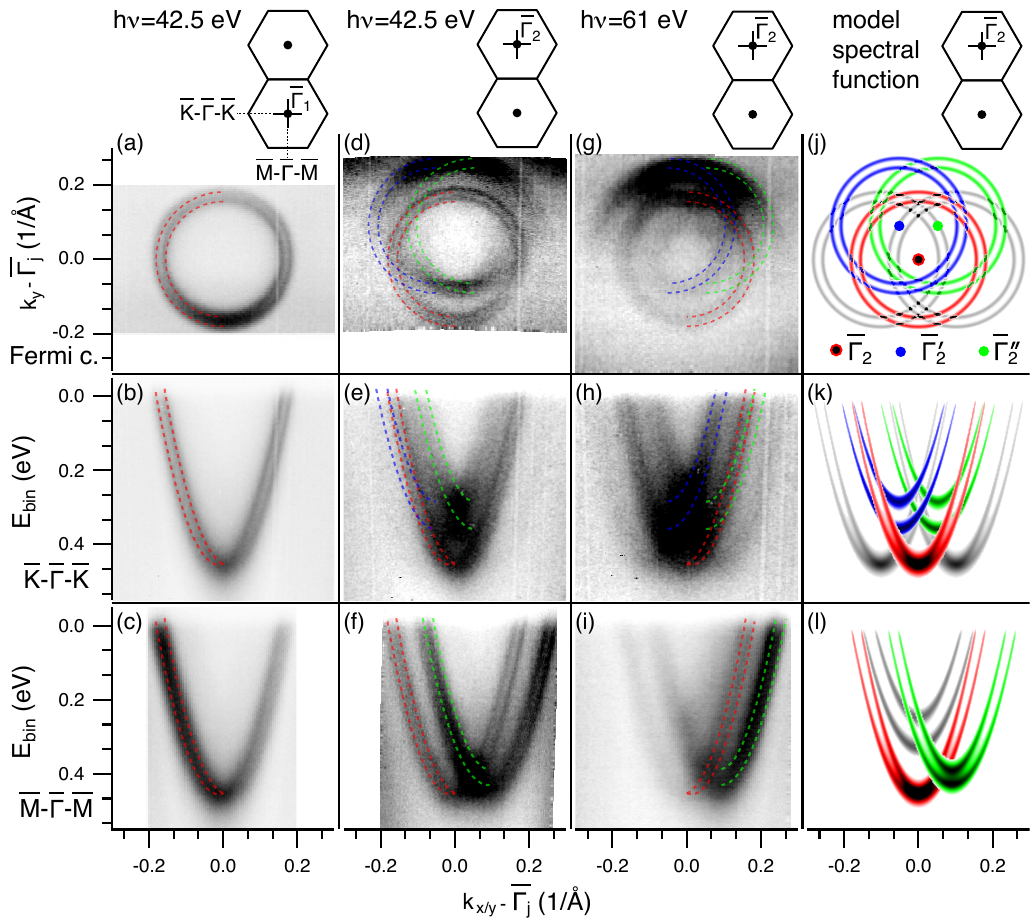}\\
\caption{(Color online) Surface electronic structure of Au(111) measured at photon energies of  42.5~eV and 61.0~eV, and calculated by a model spectral function. The photoemission intensity is shown as a function of binding energy and/or crystal momentum ($k_x,k_y$) along the cuts indicated in the sketches on top of the figure, with zero fixed at $\overbar{\Gamma}_{1}$ for panels (a)-(c) or $\overbar{\Gamma}_{2}$ (d)-(i); dark corresponds to high intensity. Panels (a)-(c) show data taken at normal emission as dispersions through the SBZ center and as constant energy contour at the Fermi energy. Panels (d)-(f) and (g)-(i) show the corresponding data taken around $\overbar{\Gamma}_{2}$, the center of the adjacent SBZ. (j) -(l) Model spectral function around $\overbar{\Gamma}_{2}$ with cuts corresponding to those in panels (a)-(i). The colored branches are centered on the first order reciprocal lattice points of the three domains. The (weaker) grey branches are additional replicas. (see Fig. \ref{fig:2}). Note that the blue and green bands in (k) and the green band in (l) do not reach to the same high binding energy as the red dispersion because the cut is not taken through their center. The colored dispersions (only symmetrical halves of each pair) are also shown as dashed lines on top of the data in (a)-(i)}
  \label{fig:1}
\end{figure*}

\section{Experimental}
The Au(111) surface was cleaned by standard methods  \cite{LaShell:1996aa,Reinert:2001aa}. The cleanliness and presence of the surface reconstruction were corroborated by STM. ARPES data were taken at the
SGM-3 beamline of the synchrotron radiation facility ASTRID2 in Aarhus \cite{Hoffmann:2004aa}. The  energy and angular resolution were better than 30~meV and 0.2$^{\circ}$, respectively. The sample temperature was 100~K. The synchrotron radiation was linearly polarized in the $k_y$ plane; incident direction and electron analyzer enclosed an angle of 50$^{\circ}$.

\section{Results and Discussion}
Figure \ref{fig:1} illustrates the effect of the herringbone reconstruction when ARPES data is taken in a higher order SBZ for two photon energies, 42.5~eV and 61.0~eV. Data taken around normal emission (Fig. \ref{fig:1}(a)-(c)) show the expected free electron-like dispersion of the state with the two spin-split bands clearly discernible and a Fermi contour consisting of two concentric circles, in excellent agreement with earlier results \cite{LaShell:1996aa,Reinert:2001aa,Nechaev:2009aa}. Note that the bands are not quite as clearly resolved as in some previous publications, notably in the direction perpendicular to the slit of the electron analyzer (along $k_y$). This is mostly due to the high photon energy used here, which results in an inferior $k$-resolution. While the periodicity of the moir\'e structure in other systems has been observed to lead to weak replicas and band gap openings in ARPES data at low temperature and with low photon energies around normal emission \cite{Reinert:2004ab}, this is not observed here, presumably because these replicas are too weak. 

 Fig. \ref{fig:1}(d)-(f) and (g)-(i)  show a measurement of the electronic structure around the $\overbar{\Gamma}_{2}$ point in the adjacent SBZ where a much more complex picture emerges. The dominating features are three versions of the original dispersion centred around different points, with Fermi contours forming a trefoil knot of spin-split circles, best seen in Fig. \ref{fig:1}(d). This results deviates in subtle but important ways from the previously discussed case of replicas induced by the  moir\'e superstructure \cite{Reinert:2004ab}, where one would expect the observation of one dominating dispersion and six very weak replicas. By contrast, the trefoil knot contour can be explained by the local uniaxial compression along $\langle 1 \overbar{1} 0\rangle$ directions in three domains of the moir\'e superstructure. In fact, the observed dispersions are not ``replicas''  in the sense of features resulting from scattering by the moir\'e periodicity, but rather are the dispersions centred around the first order reciprocal lattice points of different domains. This has the interesting consequence that two of the three dispersions stem from one domain each, in contrast to the domain averaged dispersion around $\overbar{\Gamma}_{1}$. 

\begin{figure}
\includegraphics[width=0.50\textwidth]{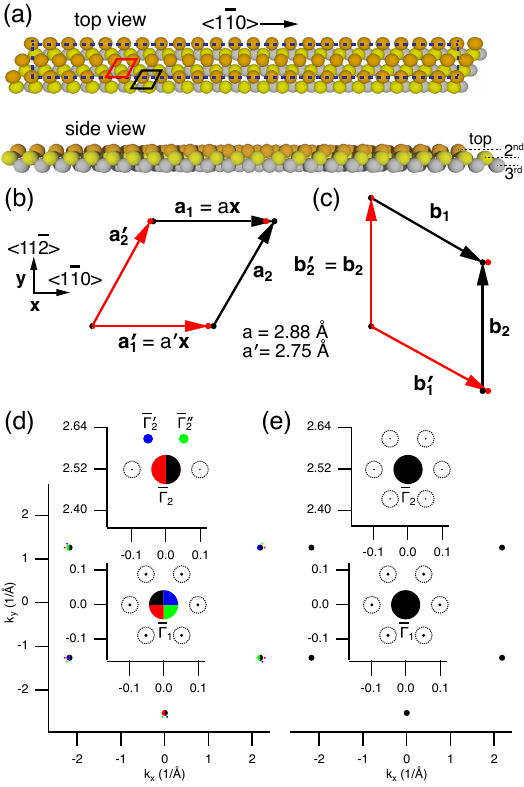}\\
\caption{(Color online) (a) Structural model of the Au(111) herringbone reconstruction, showing only one domain. The solid black and red lines give the unit cell of the truncated bulk and the first layer only, respectively. The dashed line denotes the reconstructed unit cell.  (b) Relation between the lattice vectors  $\mathbf{a}_1$, $\mathbf{a}_2$ of the truncated bulk and the lattice vectors $\mathbf{a}_{1}^\prime$ and $\mathbf{a}_{2}^\prime$ of the first layer, arising from a uniaxial compression. (c) Corresponding reciprocal lattice vectors. (d) Expected origins of surface state dispersions ($\overbar{\Gamma}_{i}$ points) for three domains with simultaneous periodicities according to the reciprocal lattices in (c). The marker radius is proportional to $|V_{\overbar{\Gamma}_{i}}|$ except for the point around $\overbar{\Gamma}_{1}$ which is scaled down by a factor of 2. Regions around the origin and the first reciprocal lattice point in the vertical direction are magnified in the insets. (e) Corresponding plot for simultaneous periodicities of the truncated bulk (black arrows in (c)) and the overall moir\'e (see Appendix). }
  \label{fig:2}
\end{figure}

The accepted structural model for one domain of the reconstruction is given in Fig. \ref{fig:2}(a) \cite{Barth:1990aa,Sandy:1991aa}. The top layer is compressed along the  $\langle 1 \overbar{1}  0\rangle$ direction, such that 23 lattice spacings of the top layer fit on 22 lattice spacings of the second layer. This results in a rectangular superstructure unit cell which is 22 times longer than the original lattice vector in one direction and $\sqrt{3}$ times this lattice vector in the other.  The key to understanding the findings in Fig. \ref{fig:1} is not this overall periodicity but the change in the local geometry required to obtain it: As shown in Fig. \ref{fig:2}(a) and (b), the compression of the first layer leads to a slight distortion of the lattice in this layer (in red) compared to the underlying crystal (in black), giving rise to an oblique lattice in contrast to the underlying hexagonal lattice. The corresponding reciprocal lattices for the first layer and the underlying lattice are given in Fig. \ref{fig:2}(c). 

Due to momentum conservation, the photoemission from solids  always involves the lattice-periodic potential and the photoemission cross section connecting a final state at wave vector $\mathbf{k}_f$ with an initial state at $\mathbf{k}_f-\mathbf{G}$ is proportional to Fourier coefficient $|V_{\mathbf{G}}|^2$ of the lattice-periodic potential  ($\mathbf{G}$ is a reciprocal lattice vector) \cite{Plummer:1982aa}. We therefore expect to observe the Au(111) surface state not only around normal emission but around all reciprocal lattice vectors $\mathbf{G}$ with a finite $V_{\mathbf{G}}$. We call these points $\overbar{\Gamma}_i$. For the limiting cases of a surface state only located in the first layer (deeper layers), these $\overbar{\Gamma}_i$ points would correspond to the red (black) reciprocal lattice points in n Fig. \ref{fig:2}(c) and one would expect to observe dispersions centered there.

If  the surface state wave function follows both the periodicity of the truncated bulk \emph{and} that of the first layer, the $\overbar{\Gamma}_i$ points for the combined reciprocal lattice and the corresponding $|V_{\overbar{\Gamma}_i}|^2$  can be obtained using the convolution theorem of Fourier transformation \cite{Zeller:2014aa}. The $|V_{\overbar{\Gamma}_i}|^2$  can be assumed to predict the relative  intensities from the surface state dispersions. Fig. \ref{fig:2}(d) shows the position of the  $\overbar{\Gamma}_i$ points for the three rotational domains of the reconstruction (see Appendix). The area of the points corresponds to the value of $|V_{\overbar{\Gamma}_i}|^2$. The color (RGB) encodes the domain a particular $\overbar{\Gamma}_i$ stems from. Black indicates contributions from all the truncated bulk. This simple model predicts the presence of three  strong dispersions along with two weak ones. The three intense dispersions are centred on the $\overbar{\Gamma}_i$ points of the first layer reciprocal lattice for the three rotational domains on the surface  $\overbar{\Gamma}_{2}$, $\overbar{\Gamma}_{2}'$, $\overbar{\Gamma}_{2}''$. The weaker dispersions can be considered to be replicas. This prediction is in excellent qualitative agreement with the results in Fig. \ref{fig:1}. Note that the kinematic diffraction pattern for the surface would be obtained in a similar way and Fig. \ref{fig:2}(d) is very similar to electron diffraction data from Au(111) \cite{Hove:1981aa}.

The physical picture leading to the pattern in Fig. \ref{fig:2}(d)  is significantly different from that previously used to account for the replica bands around normal emission. In Ref. \cite{Reinert:2004ab}, such replicas are explained by an interaction of the surface state with the long-range moir\'e pattern. The expected $\overbar{\Gamma}_i$ points for this case can also be calculated, using the large scale moir\'e structure rather than the nearly equal competing periodicities of first and deeper layers (see Appendix). The resulting $\overbar{\Gamma}_i$ points are shown in Fig. \ref{fig:2}(e) along with their $|V_{\overbar{\Gamma}_{i}}|$.  Fig. \ref{fig:2}(d) and (e) both predict weak replica bands around $\overbar{\Gamma}_1$,  consistent with the results of Ref. \cite{Reinert:2004ab}. However, the models are distinctly different for the situation close to $\overbar{\Gamma}_2$.  For an interaction with a large-scale  moir\'e structure, one would expect to observe six replicas around \emph{every} reciprocal lattice point of the undistorted lattice and the replicas around $\overbar{\Gamma}_2$ would be even weaker than those around $\overbar{\Gamma}_1$. While the intensity of the six moir\'e-induced replicas around each original reciprocal lattice point is exactly symmetric in our simple model, this would not strictly need to be so in an ARPES experiment because of $k$-dependent matrix element variations.  

The ARPES results with only three prominent Fermi contours are clearly in much better agreement with the model in Fig. \ref{fig:2}(d), as are the contours' positions. We perform a more quantitative comparison by constructing a spectral function based on the superposition of dispersions, according to the calculated $\overbar{\Gamma}_i$'s in Fig. \ref{fig:2}(d). The parameters for the dispersion are taken from a fit to the data around normal emission. The result of this is shown in Fig. \ref{fig:1}(j)-(l) with colors chosen such that the red, blue and green dispersions in the figure correspond to the  $\overbar{\Gamma}_{2}$, $\overbar{\Gamma}_{2}'$ and $\overbar{\Gamma}_{2}''$ points in Fig. \ref{fig:2}(d) and the grey dispersions correspond to the weaker replicas.  The three strong dispersions are  superimposed on the data and good agreement is found. The weak dispersions can also be seen, especially in Fig. \ref{fig:1}(g), (h). The many bands and their overlap can make the direct comparison difficult but some of the details are better seen when comparing the model to second derivatives of the photoemission data (see Appendix). The spectral function around $\overbar{\Gamma}_{2}$ is dominated by three dispersions for all photon energies studied here. The replica bands, represented by the grey dispersions, are very weak and only barely observed for $h\nu=$61~eV.  
No dispersions at the ``bottom'' of the hexagon (i.e., closer to the $\overbar{\Gamma}_{1}$ point, as would be expected for the dispersion resulting from the presence of the moir\'e) are ever observed.
The model of Fig. \ref{fig:2}(d) also predicts that the two outer dispersions (around $\overbar{\Gamma}_{2}'$ and $\overbar{\Gamma}_{2}''$) should stem from different domains, implying that there cannot be any gap openings at the crossing points of the $\overbar{\Gamma}_{2}'$ and $\overbar{\Gamma}_{2}''$ parabolas, and this is confirmed by the data (see Appendix). 

The two scenarios in Fig. \ref{fig:2}(d) and (e) are merely limiting cases and not mutually exclusive, but the strength of the effects is very different. The competing periodicities in the first and second layer leading to the trefoil knot Fermi contour are present on a  short range: If the surface state wave functions are influenced by the first layer periodicity, this immediately  leads to a significant component in the Fourier spectrum. The effect of the overall moir\'e periodicity, on the other hand, may be weak because of the weak Fourier coefficients for this modulation, the finite number moir\'e periodicities in one domain and the finite quantum coherence of the wave function \cite{Burgi:2000aa}. On the other hand, both models predict six weak replicas around  $\overbar{\Gamma}_1$ and the results reported here are therefore consistent with those of Ref. \cite{Reinert:2004ab}. Even the observation of initial state gaps between the centre dispersion and these replicas is not in contradiction to the findings here - gap openings would only be prevented between neighboring replicas as these stem from different domains of the reconstruction.

Previous ARPES results from Au(111) have shown that the consequences of herringbone reconstruction can be detected but that the effect is quite weak \cite{Reinert:2004ab,Borghetti:2012aa}.  This was in puzzling contrast to results by STM \cite{Chen:1998aa,Burgi:2002aa} and  density functional theory (DFT) \cite{Takeuchi:1991aa,Seitsonen:2016aa} which show that the surface state wave functions (or the Kohn-Sham orbitals) are strongly influenced by the reconstruction. In particular, DFT can track the surface state wave function, showing a strong localization of the state in the first layers and a periodicity that follows both the first and the lower layers. We show here that this does also lead to a strong effect in ARPES, but only as regards the surface state in a higher Brillouin zone. The separation of the trefoil knot Fermi surfaces is expected to increase still more when going to even higher order  $\overbar{\Gamma}$ points. It is also worth noting that introducing the reconstruction in DFT shifts the surface state band at $\overbar{\Gamma}$ in binding energy from 350 to 490~meV, and thereby brings it into excellent agreement with the experimental results. When the reconstruction is lifted, for example by sulphur adsorption, this binding energy is again reduced to 376(10)~meV and the trefoil knot Fermi contour disappears (see Appendix). 

We note that the approach of calculating the maps in Fig. \ref{fig:2}(d) and (e) by just two competing periodicities is, of course, a simplification. On a local scale, STM  indicates a stronger binding of the surface state electrons in the hexagonally closed packed regions between first and second layer \cite{Chen:1998aa,Burgi:2002aa}, suggesting a complex interplay between surface state and local structure. Moreover, the compression of the atoms in the first layer is not entirely uniform \cite{Barth:1990aa,Sandy:1991aa}. However, an estimate of this imperfect periodicity's effect on the expected diffraction pattern shows that it only induces minor changes (see Appendix). We also note that the simple picture presented here does not account for final state effects. 

An interesting consequence of our findings  is that it is possible to observe the surface state dispersion in a single domain, such as around $\overbar{\Gamma}_{2}'$ and $\overbar{\Gamma}_{2}''$, instead of the average from all three domains around $\overbar{\Gamma}_{1}$ or $\overbar{\Gamma}_{2}$. This opens the possibility to test the non-parabolicity of the state. Indications of this effect have previously been detected by STM on Cu(111) and Ag(111) \cite{Burgi:2000ab}  and recently even by ARPES on Au(111) for the domain-averaged Fermi contour around $\overbar{\Gamma}_{1}$ \cite{Tusche:2015aa}. A more practical consequence relates to the importance of the Au(111) surface state for the calibration of spin detectors. Such a calibration is routinely done using the surface state around $\overbar{\Gamma}_{1}$ \cite{Berntsen:2010,Takayama:2015} but there might be advantages to using the dispersion of a single domain instead. 

\section{Conclusion}
In conclusion,  trefoil knot-like constant energy surfaces  have been found for the  Au(111) surface state when performing ARPES in a higher order SBZ, solving the long-standing discrepancy between, on one hand, the strong effect of the reconstruction observed in STM \cite{Chen:1998aa,Burgi:2002aa} and DFT \cite{Takeuchi:1991aa,Seitsonen:2016aa} and, on the other hand, its near absence in ARPES data. The approach of probing higher order SBZ should be generally applicable to probe similar effects of strain in interfaces between two-dimensional materials and other moir\'e structures.  

\section{Acknowledgement}

We gratefully acknowledge funding from the VILLUM FONDEN via the Centre of Excellence for Dirac Materials (Grant No. 11744), the Aarhus University Research Foundation and the Danish Council for Independent Research, Natural Sciences under the Sapere Aude program (Grant No. DFF-4002-00029). We thank Arlette S. Ngankeu and Jill A. Miwa for experimental help and discussions.

\section{Appendix}

This Appendix contains an evaluation of the expected positions of additional surface state dispersions due to the presence of the herringbone reconstruction, an estimate of the effect of the top layer's non-uniform compression on the results, a detailed comparison between our model and the second derivative of the ARPES data, a demonstration of increased photoemission intensity at the crossing points of dispersions and photoemission results for a lifted herringbone reconstruction.

%
%
\begin{section}{A. Expected Origins of Surface State Dispersions}

In order to illustrate the different effects of the local strain and the overall moir\'e periodicity on the expected position of surface state dispersion origins ($\Gamma_{i}$ points), we first consider a one-dimensional model in Fig. \ref{fig:s1}. The figure and the discussion are adapted from the recent work on moir\'e structures of graphene by Zeller and G\"unther in Ref. \onlinecite{Zeller:2014aa}. We consider two periodic functions $f_1(x)$ and $f_2(x)$, both of the form $f_i(x) = 1 + \cos(k_i x)$ (the actual values of the Fourier coefficients could be different for the two functions but this does not matter for the simple illustration here). The overall periodicity of the system shall be given by the product of $f_1$ and $f_2$. Fig. \ref{fig:s1}(a) shows this situation represented by the Fourier coefficients $F\{f_1\}$ and $F\{f_2\}$, describing $f_1$ and $f_2$, such that $f_1$ represents the periodicity of the Au(111) truncated bulk lattice parallel to the surface and $f_2$ represents the long-range moir\'e periodicity periodicity (with a small $k_2$). If only one periodicity were present (either only the lattice or only the moir\'e), the location of non-zero Fourier coefficients would give the $\Gamma_{i}$ points. When seeking the $\Gamma_{i}$ points for the product, we need to evaluate 
\begin{multline}
F\{f_1 \times f_2\} (k) = F\{f_1\}(k) \otimes F\{f_2\}(k) \\ = \int F\{f_1\}(k-k')  F\{f_2\}(k') dk'
\end{multline}
where $\otimes$ denotes the convolution of $F\{f_1\}$ and $F\{f_2\}$. This is done graphically: Finite contributions for $F\{f_1\} \otimes F\{f_2\}$ arise when $F\{f_1\}$ is shifted by a $k$ value such that finite Fourier coefficients in $F\{f_1\}$ come into registry with finite coefficients in $F\{f_2\}$. This is the case for shifts by the colored arrows in Fig. \ref{fig:s1}(a). The resulting values of $F\{f_1\} \otimes F\{f_2\}$ can also be directly read from the figure by multiplying the lengths of the overlapping bars: For the orange arrow it is $1 \times 0.5= 0.5$ and for the blue it is $0.5 \times 0.5= 0.25$. $F\{f_1\} \otimes F\{f_2\}$ now gives the $\Gamma_{i}$ points for the combined system and, at this simple level of theory, the value of the Fourier coefficients represents the strengths of the expected dispersions. Replicas would be found around the origin and around the higher order diffraction spots of the regular lattice and the latter replicas would be weaker than the former. 

Fig. \ref{fig:s1}(b) shows the same construction for a situation dominated by the simultaneous presence of the two local periodicities. $f_1$ again represents the periodicity of the Au(111) truncated bulk lattice parallel to the surface but $f_2$ now corresponds to the slightly compressed periodicity of the top layer. Following the identical construction of $F\{f_1\} \otimes F\{f_2\}$, we obtain only two $\Gamma_i$ values in the vicinity of the original single higher order spot of the lattice, not three. Moreover, the Fourier coefficients of the two have equal values in this model.

\begin{figure}[!h]
	\includegraphics[width=0.5\textwidth]{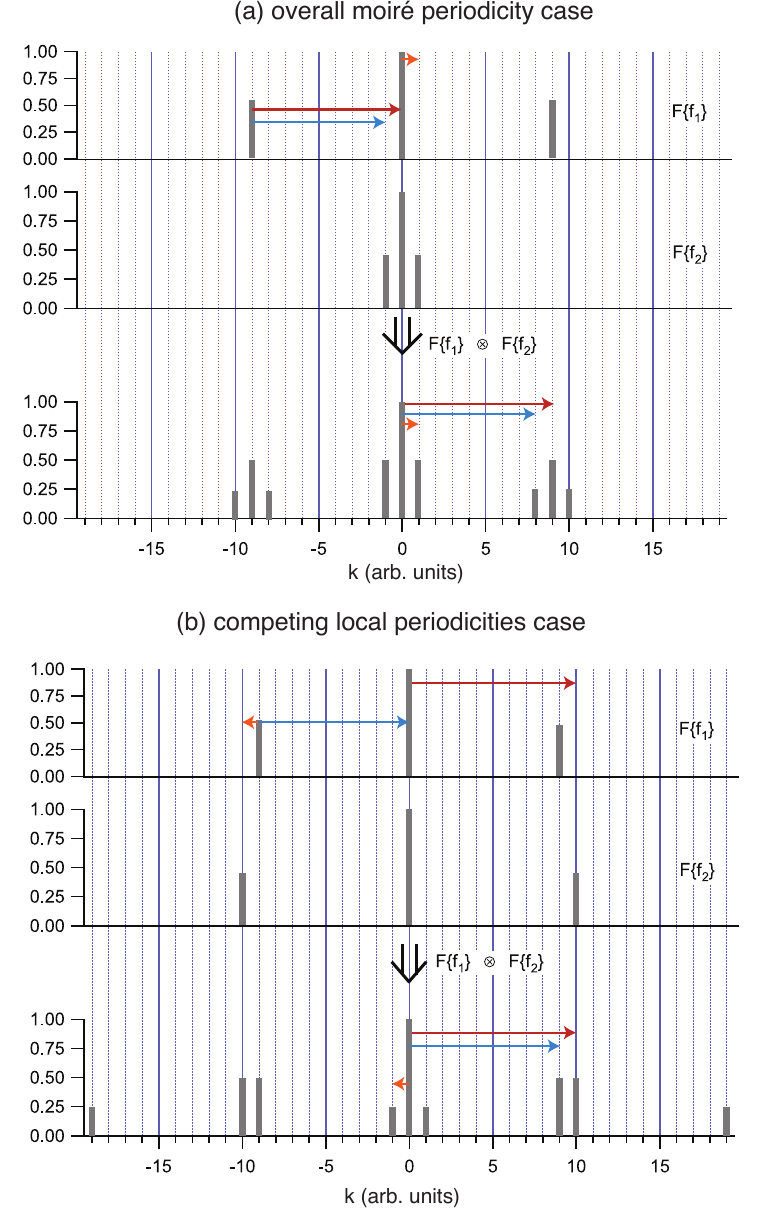}\\
	\caption{One-dimensional model for the calculation of expected surface state dispersion origins. (a) Situation for the simultaneous presence of truncated bulk lattice and  overall moir\'e periodicity. (b) Situation for simultaneous presence of the  truncated bulk lattice and the slightly shorter periodicity in the top layer. The arrows illustrate the convolution of the two functions. Each colored arrow corresponds to a $k$ movement that brings finite Fourier components in $F\{f_1\}$ in registry with those in $F\{f_2\}$ and thereby leads to finite values in $F\{f_1\} \otimes F\{f_2\}$ at this $k$. }
	\label{fig:s1}
\end{figure} 

Before extending this model to two dimensions, we briefly give explicit expressions of the top layer and projected bulk lattices:

\begin{equation}
\mathbf{a}_{1}=a(1,0), \mathbf{a}_{2}=a\left(\frac{1}{2},\frac{\sqrt{3}}{2}\right),
\end{equation}

\begin{equation}
\mathbf{a}_{1}^\prime=a^\prime(1,0), \mathbf{a}_{2}^\prime=\left(\frac{a^\prime}{2},\frac{a\sqrt{3}}{2}\right),
\end{equation}
where $a$=2.88 \AA~ and $a^\prime$=2.75 \AA. Thus, corresponding reciprocal lattices are given by:

\begin{equation}
\mathbf{b}_{1}=2\pi\left(\frac{1}{a},-\frac{1}{a\sqrt{3}}\right), \mathbf{b}_{2}=2\pi\left(0,\frac{2}{a\sqrt{3}}\right),
\end{equation}

\begin{equation}
\mathbf{b}_{1}^\prime=2\pi\left(\frac{1}{a^\prime},-\frac{1}{a\sqrt{3}}\right), \mathbf{b}_{2}^\prime=2\pi\left(0,\frac{2}{a\sqrt{3}}\right).
\end{equation}
Three-fold symmetry of the underlying bulk is reflected by the existence of rotational domains of the top layer which are taken into account in the model by $2\pi/3$~and $4\pi/3$ rotated $\mathbf{b}_{\mathrm{i}}^\prime$.

With this, we numerically calculate the equivalent of  Fig. \ref{fig:s1} in  two dimensions in Figs. \ref{fig:s2a} and  \ref{fig:s2b} and we present these results schematically in Fig. 2(d) and (e) of the main paper. As in the one-dimensional model, the choice of the Fourier coefficients $V_{\mathbf{G}}$ in $f_1(\mathbf{k})$ and $f_2(\mathbf{k})$ is arbitrary.  Fig. \ref{fig:s2a} shows the evaluation of  the $\overbar{\Gamma}_i$ and $|V_{\overbar{\Gamma}_i}|$ for the interaction of the truncated bulk lattice (periodicity $\mathbf{b}_{1}$ and $\mathbf{b}_{2}$, Fig. \ref{fig:s2a}(a)) with the moir\'e periodicity in one direction (Fig. \ref{fig:s2a}(b)). As in Ref. \onlinecite{Reinert:2004ab}, the moir\'e periodicity is modelled by only one Fourier component in the long direction of the moir\'e (22 times the lattice constant) and none in the short direction ($\sqrt{3}$ times the lattice constant). The convolution of the two images of Fig. \ref{fig:s2a}(a) and (b) is shown in Fig. \ref{fig:s2a}(c). It leads to the presence of two weak spots around every position of non-zero Fourier coefficient of the lattice. These spots are weaker around higher order lattice points than around the origin. Fig. \ref{fig:s2a}(d) shows the summation of the result in (c) for all three domains present on the surface. The result is consistent with the experimental findings in Ref. \onlinecite{Reinert:2004ab}.

\begin{figure}[!h]
	\includegraphics[width=0.45\textwidth]{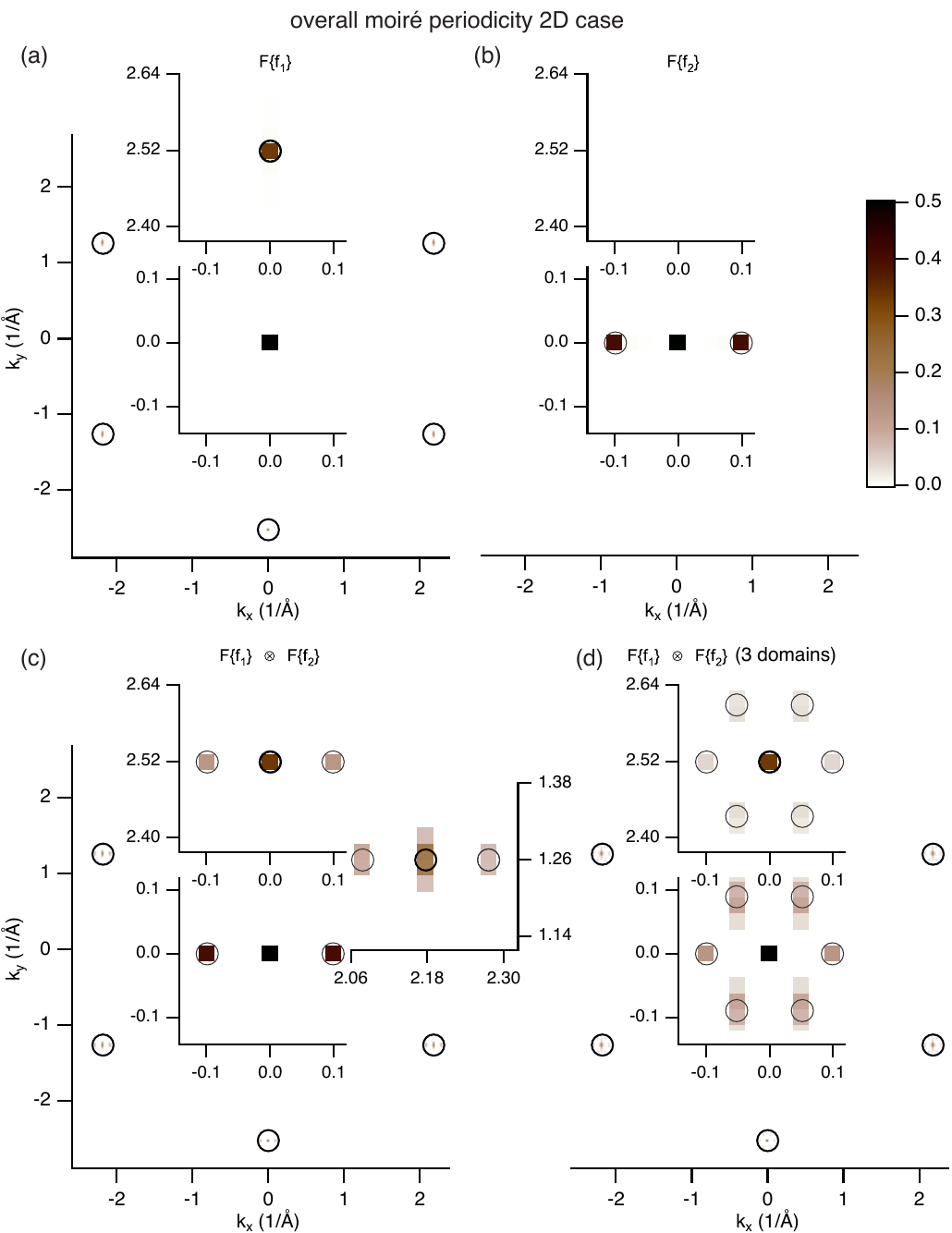}\\
	\caption{(a) and (b) Fourier coefficients $|V_{\mathbf{G}}|$ for the truncated bulk lattice ($\mathbf{b}_{1}$ and $\mathbf{b}_{2}$) and the moir\'e periodicity in one domain, respectively. The intensity is normalized to the zeroth order component and mapped with the colors scale included in the figure. The peaks are, in some cases, difficult to see and therefore partly marked by circles around them. The regions around the origin and the first reciprocal lattice point in the vertical direction are magnified in the insets. (c) Convolution of (a) and (b). (d) Sum of convolution of (a) and (b) for the three possible domains on the surface.}
	\label{fig:s2a}
\end{figure} 

The corresponding situation for the simultaneous presence of the truncated bulk  ($\mathbf{b}_{1}$ and $\mathbf{b}_{2}$)  and top layer ($\mathbf{b}_{1}^\prime$  and $\mathbf{b}_{2}^\prime$) periodicities is shown in Fig. \ref{fig:s2b}. If we choose the domain of interest in the same way as in Fig. 2 of the main paper, the reciprocal vectors $\mathbf{b}_{2}$ and $\mathbf{b}_{2}^\prime$ are identical and therefore only one intense spot appears around the position called $\overbar{\Gamma}_2$ in Fig. 2 of the main paper. However, the convolution also gives rise to two weak replicas close to this. For other first order lattice spots (see second inset in Fig. \ref{fig:s2b}(c)), the first order reciprocal lattice vectors are not the same for the top layer and truncated bulk and thus two strong Fourier components appear but no additional weak ones. Finally, when summing over all three domains, every first order reciprocal lattice vector shows three high coefficients in its vicinity and two weak ones.  (Fig. \ref{fig:s2b}(d)). Figs. 2(d) and (e) show a schematic representation of these results in which the magnitudes of the Fourier coefficients are encoded in the size of the markers. 

\begin{figure}[!h]
	\includegraphics[width=0.45\textwidth]{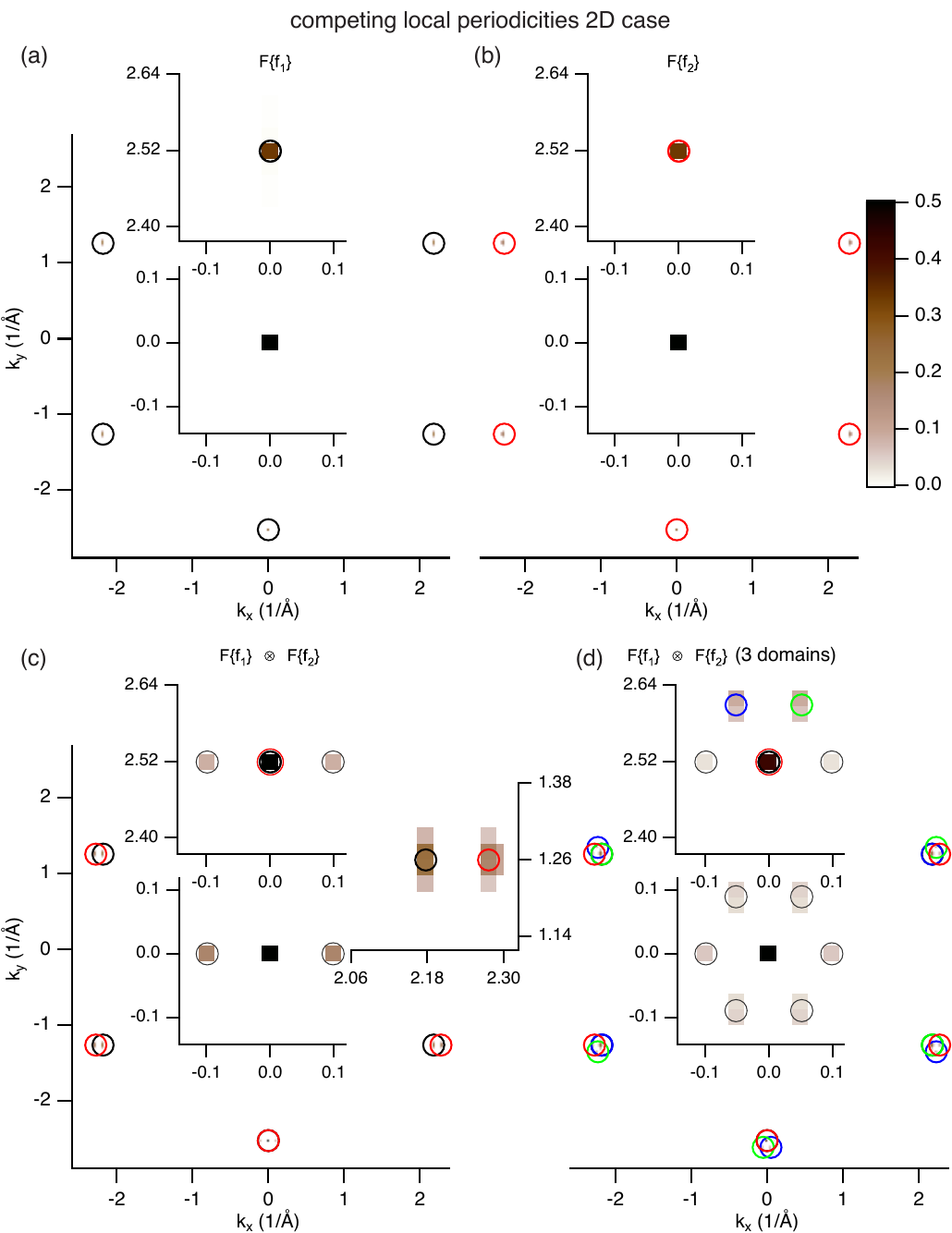}\\
	\caption{(a) and (b) Fourier coefficients for the truncated bulk lattice ($\mathbf{b}_{1}$ and $\mathbf{b}_{2}$) and the surface layer lattice ($\mathbf{b}_{1}^\prime$  and $\mathbf{b}_{2}^\prime$), respectively. The intensity is normalized to the zeroth order component and mapped with the colors scale included in the figure. The peaks are very narrow and therefore partly marked by circles around them. The regions around the origin and the first reciprocal lattice point in the vertical direction are magnified in the insets. (c) Convolution of (a) and (b). Here insets show magnifications around the origin and around two first order lattice spots. For this particular domain orientation, one strong Fourier coefficients and two weak ones are found close to one of these points and two strong ones close to the other.  (d) Sum of convolution of (a) and (b) for the three possible domains on the surface.}
	\label{fig:s2b}
\end{figure} 

The compression of the atoms in the top layer is not entirely uniform \cite{Barth:1990aa,Sandy:1991aa} and it is interesting to ask how this affects the picture given above. We test this in the one-dimensional model in Fig. \ref{fig:s3}. Fig. \ref{fig:s3}(a) juxtaposes a lattice manifesting the imperfect, position-dependent periodicity previously discussed in Refs. \onlinecite{Barth:1990aa,Sandy:1991aa} (red line, with changing periodicity indicated quantitatively at top of panel) with a perfectly periodic function (black dashed line). Fig. \ref{fig:s3}(b) shows the corresponding Fourier transformations. The positions of the first order Fourier components are the same for both cases. The imperfect periodicity leads to a loss of intensity in the first order Fourier coefficients and to very broad side bands around their positions. The consequence of this for the surface state dispersion is a weakening of the coherent dispersions and a slight increase of the background. While the distortion is much bigger than typical atomic vibrations, the result is thus similar to what is described by a Debye-Waller factor in X-ray diffraction. 

\begin{figure}[!h]
	\includegraphics[width=0.45\textwidth]{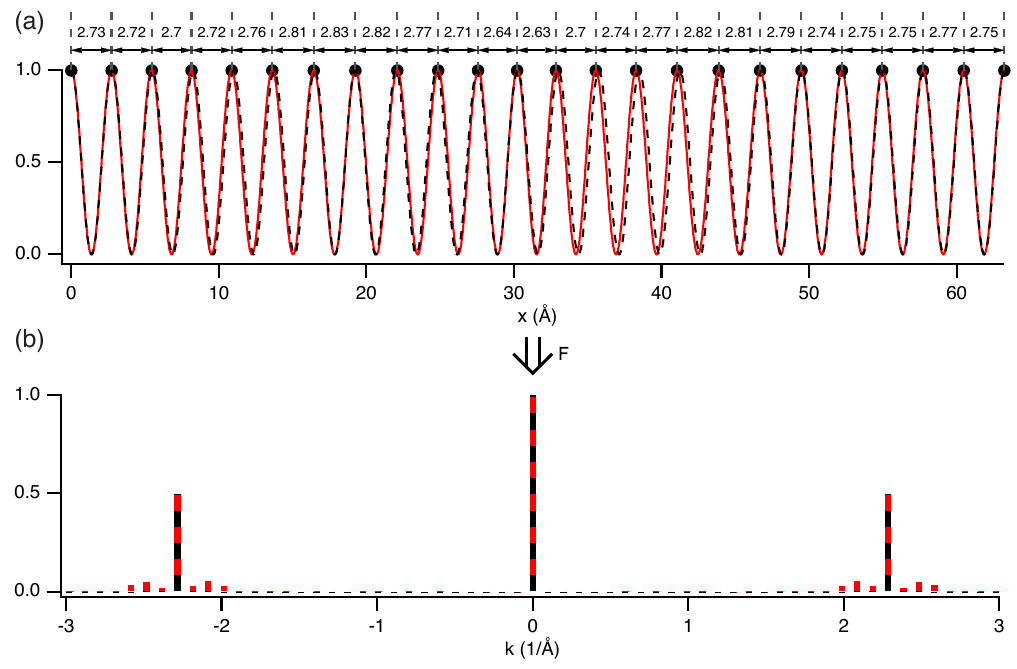}\\
	\caption{Effect of an irregular lattice spacing in the top layer. (a) One-dimensional model of the non-uniform compression in the top layer with the interatomic distances given in $\AA$. The black dashed line is a periodic function for the perfect lattice. The red line is a quasi-periodic function tracking the position of the top layer atoms. (b) Fourier transformation of the two functions.  }
	\label{fig:s3}
\end{figure} 
\end{section}
\vspace{1mm}

 \begin{section}{B. Quantitative comparison between data and model}
 
 Fig. \ref{fig:s4} is the same as Fig. 1 of the main paper, except that it shows the second derivative of the photoemission spectra. This approach is frequently used to show weak features on a high background. While it has some drawbacks (shifts of the maxima and difficulties in the presence of many peaks), it nevertheless helps to disentangle the dispersions here.  Fig. \ref{fig:s5} is the same but without the calculated dispersions superimposed on the data. Overall, a good agreement of model and data is found.

 Note that the model is somewhat oversimplified.  For example, the colored dispersions here were constructed to be the same as the one around $\overbar{\Gamma}_{1}$, and this does not necessarily need to be the case in the real physical system. The dispersion around $\overbar{\Gamma}_{1}$  is an average over all domains while the colored dispersions are (partly) from single domains. 
 
\begin{figure*}[!h]
	\includegraphics[width=0.8\textwidth]{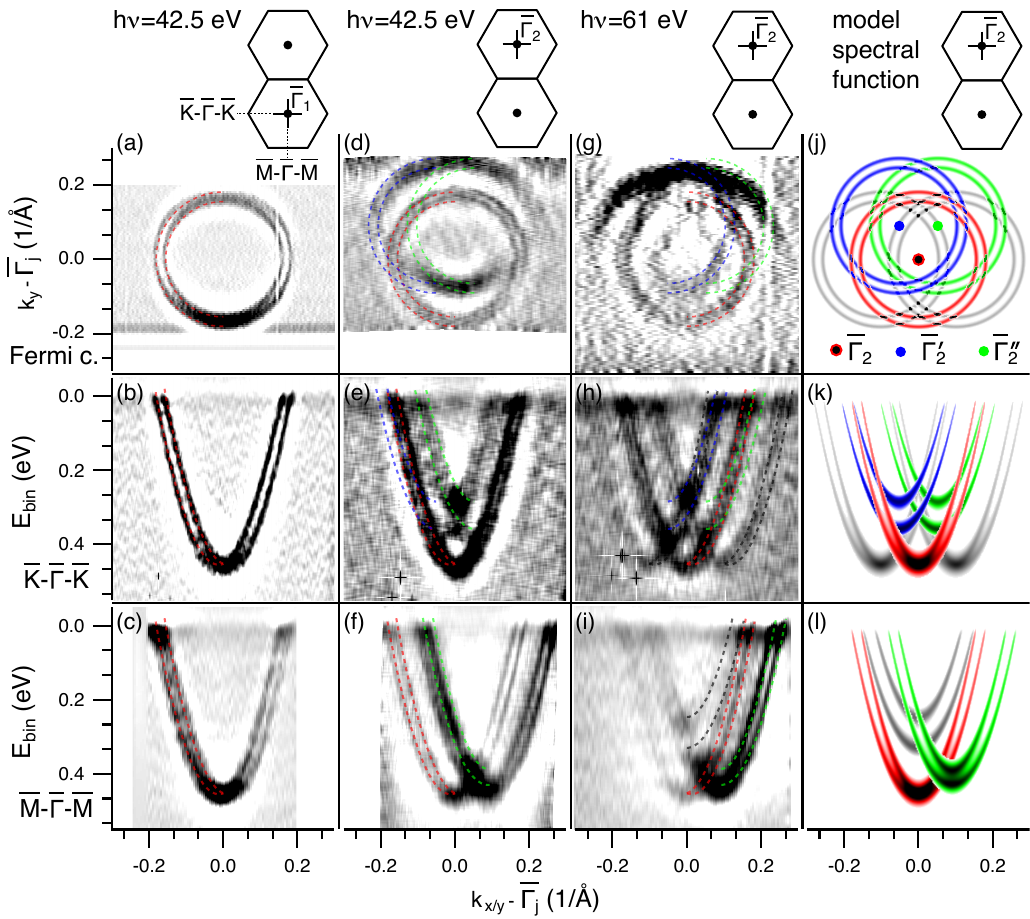}\\
	\caption{Same figure as Fig. 1 in the main paper, but showing second derivatives of the photoemission intensity instead of the intensity as such. The second derivative is shown as a function of binding energy and/or crystal momentum ($k_x,k_y$) along the cuts indicated in the sketches on top of the figure, with zero fixed at $\overbar{\Gamma}_{1}$ for panels (a)-(c) or $\overbar{\Gamma}_{2}$ (d)-(i). Panels (a)-(c) show data taken at normal emission as dispersions through the SBZ center and as a constant energy contour at the Fermi energy. Panels (d)-(f) and (g)-(i) show the corresponding data taken around $\overbar{\Gamma}_{2}$, the center of the adjacent SBZ. (j) -(l) Model spectral function around $\overbar{\Gamma}_{2}$ with cuts corresponding to those in panels (a)-(i). The colored branches are centered on the first order reciprocal lattice points of the three domains. The (weaker) grey branches are additional replicas. The colored dispersions are also shown as dashed lines on top of the data in (a)-(i). Note that the blue and green bands in (k) and the green band in (l) do not reach to the same high binding energy as the red dispersion because the cut is not taken through their center.}
	\label{fig:s4}
\end{figure*}

\begin{figure*}[!h]
	\includegraphics[width=0.8\textwidth]{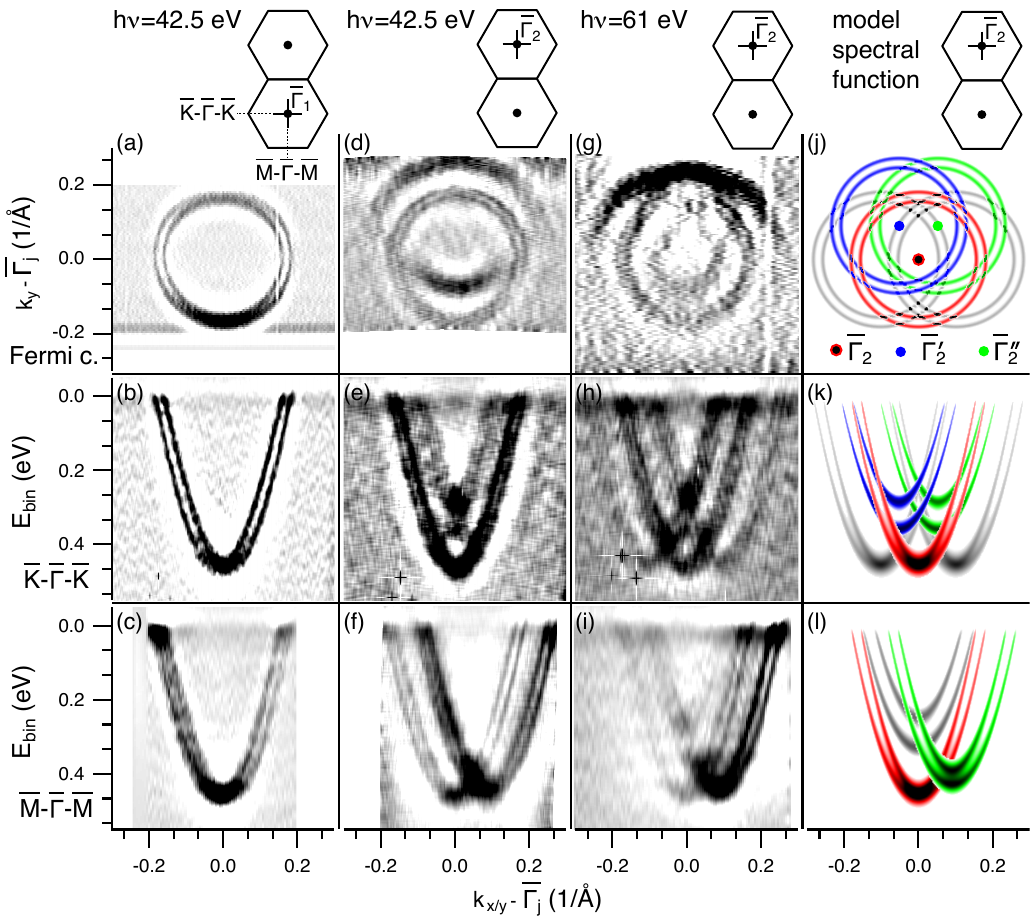}\\
	\caption{Same as Fig. \ref{fig:s4} but without the calculated dispersions superimposed on the data.}
	\label{fig:s5}
\end{figure*}

The expected effect of increased photoemission upon the crossing of bands from different domains is seen at several points in the data of the previous figures, and Fig. \ref{fig:s6} gives another good illustration of this. When the electronic wave function is modulated by a large wavelength envelope function, one can expect gap openings between the original band and the replicas. This was indeed observed for Au vicinal surfaces \cite{Burgi:2002aa} and graphene/Ir \cite{Pletikosic:2009aa}. In the overall moir\'e periodicity picture one would expect to observe the same behavior around $\overbar{\Gamma}_{2}$, i.e., gap opening along the potential periodicity direction $\overbar{\Gamma}$-$\overbar{K}$. Here, however, exactly the opposite is observed:  the intensity is enhanced at the crossing (see Fig.~\ref{fig:s6}), a result which can only be explained by the competing local periodicities picture. In this framework the surface states originate from different domains on the sample and, due to the large area probed with ARPES, the signal from all domains is incoherently added. 

\begin{figure}[!h]
	\includegraphics[width=0.4\textwidth]{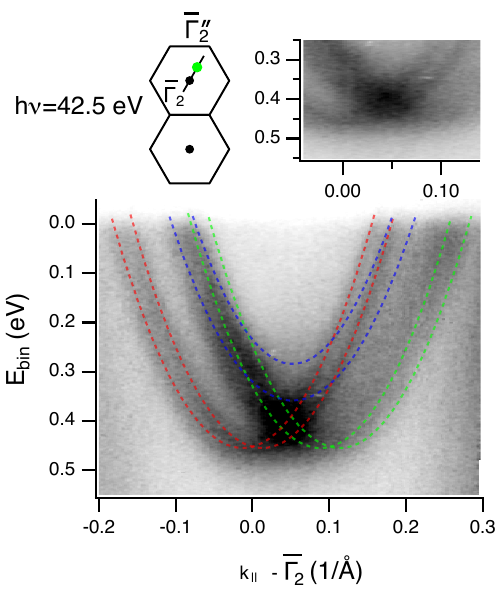}\\
	\caption{(Color online) Surface electronic structure of Au(111) along the cut $\overbar{\Gamma}_{2}$-$\overbar{\Gamma}_{2}''$ ($\overbar{\Gamma}$-$\overbar{K}$), as indicated in the sketch on top of the figure, measured  at photon energy of  42.5~eV. The photoemission intensity is shown as a function of binding energy and crystal momentum $k_{\parallel}$ with zero fixed at $\overbar{\Gamma}_{2}$. The colored dispersions are shown as dashed lines on top of the data. The inset is a magnification of the same data and an intensity enhancement, i.e. no gap opening, at the crossing of surface states originating from different rotational domains. }
	\label{fig:s6}
\end{figure}

\end{section}

\begin{section}{C. Corresponding results with a lifted reconstruction}

Fig. \ref{fig:s7} illustrates that the effect of replicas vanishes when the herringbone reconstruction is lifted by exposure to H$_2$S. This is  presumably because H$_2$S exposure induces the adsorption of small quantities of sulphur. In order to lift the herringbone reconstruction, the sample was  annealed (up to ca. 350$^{\circ}$C) for 20 min in  H$_2$S  (at $5\cdot10^{-5}$~mbar). A Lorenzian fit to the energy distribution curve through  $\overbar{\Gamma}$ gives a maximum binding energy of $E_0=376(10)$ and $E_0=453(3)$~meV for the unreconstructed and reconstructed surfaces, respectively. 

\begin{figure*}[!h]
	\includegraphics{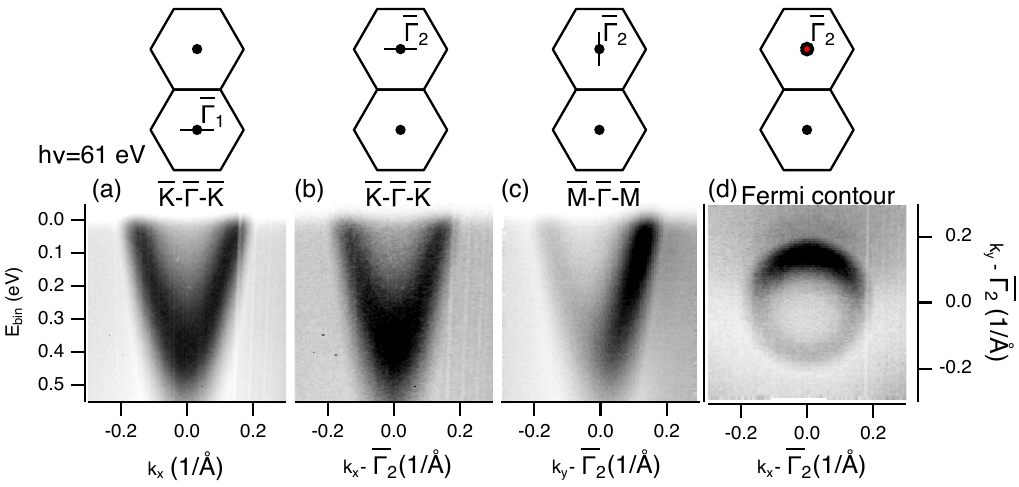}\\
	\caption{Surface electronic structure of Au(111) after lifting the herringbone reconstruction by exposure to H$_2$S, measured at a photon energy of 61.0~eV. The photoemission intensity is shown as a function of binding energy and / or crystal momentum ($k_x,k_y$) along the cuts indicated in insets with zero fixed at $\overbar{\Gamma}_{1}$ for panel (a) and $\overbar{\Gamma}_{2}$ for panels (b)-(d); dark corresponds to high intensity. }
	\label{fig:s7}
\end{figure*}

\end{section}

%
\end{document}